\documentclass[a4paper,10pt]{article}
\usepackage[utf8x]{inputenc}

\usepackage{fullpage}
\usepackage[pdftex]{graphicx}
\usepackage{graphics}
\usepackage{amsmath}
\usepackage{amssymb}
\usepackage{txfonts}
\usepackage{textcomp}
\usepackage{amsfonts}
\usepackage{amssymb}

\title{\bf{General Scattering Mechanism and Transport in Graphene}}
\author{M. Rabiu\thanks{E-mail adress: rabpeace10gh@gmail.com}\,\,\,$^a$, S. Y. Mensah$^b$ and S. S. Abukari$^c$}
\date{}
\begin{document}
\maketitle
\begin{center}
\emph{$^{\bf a}$University for Development Studies, Faculty of Applied Science, Department of Applied Physics, Navrongo, Ghana \\
$^{\bf b}$University of Cape Coast, Center for Laser and Fibre Optics, Physics Department, Cape coast, Ghana\\
$^{\bf c}$University of Cape Coast, School of Physical Sciences, Department of Physics, Cape Coast, Ghana}
\end{center}

\begin{abstract}
Using quasi-time dependent semi-classical transport theory in RTA, we obtained coupled current equations in the presence of time varying field and based on general scattering mechanism $\tau \propto \mathcal{E}^{\beta}$. We find that close to the Dirac point, the characteristic exponent $\beta = +2$ corresponds to acoustic phonon scattering. $\beta = +1$ long-range Coulomb scattering mechanism. $\beta = -1$ is short-range delta potential scattering in which the conductivity is constant of temperature. The $\beta = 0$ case is ballistic limit. In the low energy dynamics of Dirac electrons in graphene, the effect of the time-dependent electric field is to alter just the electron charge by $e \to e(1 + (\Omega \tau)^2)$ making electronic conductivity non-linear. The effect of magnetic filed is also considered.
\end{abstract}


\subsection*{I. Introduction}
Quite recently, semiconductor nanostructures have become the model of choice for investigation of electrical conduction. The unique two-dimensional material graphene which was first thought to be an academic material is not an exception. This 2D nanomaterial is fast becoming better candidate for electronic devices. Not only because of its noble electronic transport properties \cite{AHCastro}, but it also promises a good future in graphene based electronics industry. Graphene has received a wide academic attention and serve as a bridge between condensed matter physics and high energy physics \cite{MIKatsenelson}. The advantage of this planner material is that, one can easily change its electronic properties by introducing tunable gap in the sample \cite{LCi} or changing the number of graphene planes \cite{FGuinea}. It is also possible to fabricate free standing graphene sheets \cite{SShivaraman}. Intrisic superconducting states can also be realized in graphene \cite{BUchoa}. These among other things, means that the electronic properties of graphene can easily be tailored to fit device conditions.

The crystal structure of graphene is made up of mono layer of carbon atoms arranged in hexagonal lattice. The low energy dynamics of fermions in graphene is characterized by linear dispersion, $\mathcal{E}(k) \sim v_F|p|$. In a cleaned sample, the conduction and valence bands touch at two inequivalent Dirac points located at the corners of Brillouin zone. To understand the low energy transport in graphene, a relevant scattering potential (mechanism) is essential. However, as we shall see in this report, one does not need to consider any explicit form of scattering potential. It was shown in \cite{SAdam} that Boltzmann theory with long-range coulomb scattering can account for all experimental findings. Especially, when the electronic density around Dirac points are normalized. Also, within the Boltzmann theory using random phase approximation, coulomb scattering has been predicted to be the dominant scattering mechanism \cite{SAdam1}. Several theories including Boltzmann Transport Equation (BTE) suggest a non-universal behavior of minimal conductivity which nonetheless coincides with experimentally observed value times, i.e $\pi$, $\sigma_{theory} = \sigma_{exp}/\pi$ \cite{AKGeim, NMRPeres}. The same BTE predict other transport coefficients which agrees well with experiment \cite{NMRPeres, TStauber, JNilson}.

In this brief report, we reproduce transport properties of graphene. Within the BTE and with energy dependent relaxation time depending on power law, we showed dependence of these transport coefficients on frequency of the applied field. The remaining of this paper is organized as follows; Section {\bf II} formulates BTE and provides arguments leading to a quasi time-dependent solution (t-BTE). In section {\bf III}, we used the t-BTE to derive coupled current equations from which we derived conductivity and other transport quantities. The conductivity tensor is rederived in the presence of magnetic field in {\bf IV}. The last section {\bf V} contains discussion, conclusion and recommendations.

\subsection*{II. BTE and quasi time-dependent solution}
A time time-dependent linearized BTE has the following form \cite{JMZiman}
\begin{equation}
 \partial_tf(p) + v(p)\cdot\left[ -\frac{\mathcal{E} - \mu}{T}\nabla_rT + e(\,E(t) - \nabla_r(\mu/e)\,)\right]\left(-\frac{\partial f(p)}{\partial\mathcal{E}}\right) = \mathcal{C}f(p)\label{eq:one}
\end{equation}
where $f(p)$ depends on $t$, $r$ and $p$, i.e \[f(p) \equiv f(p,r,t).\] The group velocity, $v$ is constant of time, $\mathcal{C}f$ is the scattering term and $T$ is the lattice temperature. The applied electric field has the form $E(t) =  Ecos\Omega t$. $H$ is the magnetic field. Exact analytical solutions of eq.\eqref{eq:one} are very difficult to obtain. Especially, due to the non-linearity of $\mathcal{C}f$ and the fact that $v$ can depend on time in general. In view of this, we adopt some approximations including relaxation time approximation where
\begin{equation}
 \mathcal{C}f(p) = -\Gamma (\mathcal{E})(f(p) - f_0(p)) \label{eq:two}
\end{equation}
$\Gamma$ is inverse of relaxation time $\tau$. $f$ and $f_0$ are the time-dependent and time-independent Fermi-Dirac distribution functions. Motivated by \cite{SYMensah} in the absence of magnetic field ($B=0$), we consider a picture where the only time dependent quantity in eq.\eqref{eq:one} is the electric field. Note that Mensah solution considered the space term as perturbation. Under the above simplified assumptions together with the steady state solution \cite{JMZiman},\cite{OtfriedMadelung} the quasi solution is
\begin{equation}
 f(p) = \Gamma\int_0^{\infty} e^{-\Gamma t'}f_0(p) \,dt - e\int_0^{\infty} e^{-\Gamma t'}v(p)\cdot\left[ \frac{\mathcal{E} - \mu}{eT}\nabla_rT + (\,\nabla_r(\mu/e) - E(t')\,)\right]dt\left(-\frac{\partial f(p)}{\partial\mathcal{E}}\right), \label{eq:twoa}
\end{equation}
so that it can easily reduce back to \cite{JMZiman} when $\Omega = 0$.

\subsection*{III. Coupled currents and transport coefficients}
The sheet current for electron and energy flux in graphene are defined as \cite{MLundstrom} \cite{MRabiu}
\begin{equation}
 J_e = \frac{g_sg_ve}{A}\sum_{p}v(p)f(p) \label{eq:three}
\end{equation}
and
\begin{equation}
 J_{\mathcal{E}} = \frac{g_sg_v}{A}\sum_{p}\mathcal{E}(p)f(p), \label{eq:four}
\end{equation}
where $g_s$, $g_v$ are spin and valley degeneracies and $A$ is the graphene sheet area. We convert the sums in eqs.\eqref{eq:three} and \eqref{eq:four} to integrals following \cite{MLundstrom} 
\[ \sum_{p} \to \frac{A}{(2\pi\hbar)^2}\int_{-\pi}^{\pi}d\theta\int_{0}^{\infty}pdp,\]
substituting eq.\eqref{eq:twoa} in to eqs.\eqref{eq:three}, \eqref{eq:four} and simplifying using an energy dependent relaxation time 
\begin{equation}
 \tau(\mathcal{E}) = \Lambda \mathcal{E}^{\beta}.
\end{equation}
$\Lambda$ is constant of energy with dimensions of $s/J^{\beta}$ and $\beta$ is characteristic exponent which determines the specific type of scattering mechanism involved. The coupled current equations are 
\begin{subequations}\label{eq:five}
\begin{align}
 \label{eq:fiveA}
  J_e = \sigma(\Omega)(-\nabla_r \Phi) + S(-\nabla_r T)\\ \nonumber\\
  J_{\mathcal{E}} = TS(-\nabla_r\Phi) +  K(-\nabla_rT)
\end{align}
\end{subequations}
The measured electrochemical potential gradient is $\nabla \Phi = \nabla_r (\mu/e) - E$. 
The coefficients in eqs.\eqref{eq:five} are;
\begin{eqnarray}
  \sigma_{\beta}(\Omega) &=& \frac{ \sigma_{min}}{u_0^2}\frac{\mu^{\beta + 1} + \dfrac{\pi^2}{6}(k_BT)^2\beta(\beta + 1)\mu^{\beta - 1} + \cdots}{1 - i\Omega\Lambda\mu}. \label{eq:sigma1}\\ \nonumber\\
  S_{\beta} &=& \frac{\sigma_{min}}{u_0^2}\frac{\pi^2}{3eT}(k_BT)^2(\beta + 1)\mu^{\beta}  + \cdots .\\ \nonumber\\
  K_{\beta} &=& \frac{\sigma_{min}}{u_0^2}\frac{\pi^2}{3eT}(k_BT)^2\mu^{\beta + 1}  + \cdots .
\end{eqnarray}
$\sigma_{min}$ is the minimal conductivity in graphene expressed as $\sigma_{min} = 2e^2/h$ and $u_0$ is a constant given by $u_0^2 = \hbar/\Lambda$ and has energy dimension.

\begin{figure}[h!]
 \centerline{\includegraphics[width=6.0in]{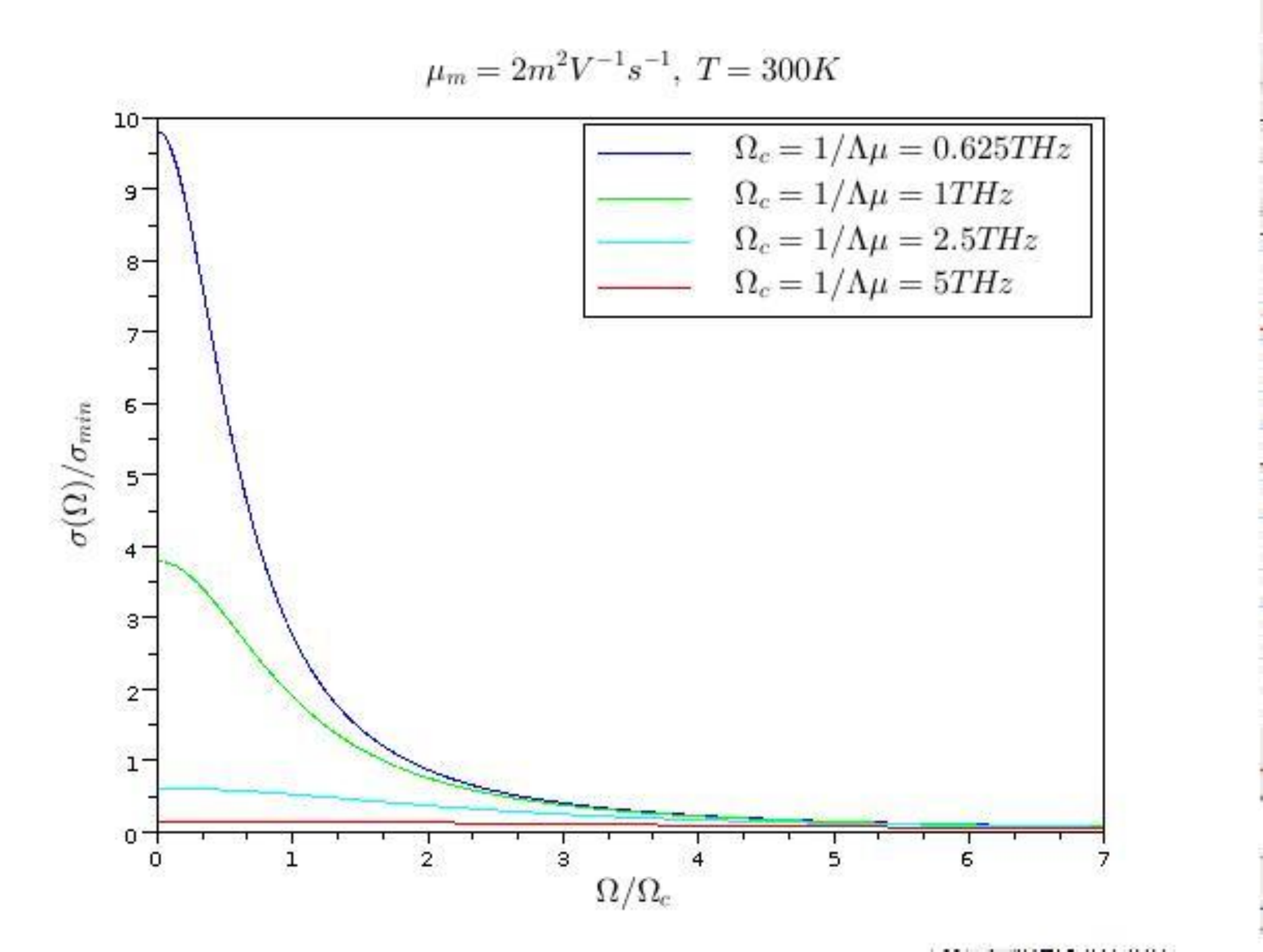}}
 \caption{Normalized conductivity is plotted against $\Omega/\Omega_c$ at fixed values of doping; $\mu = 0.8eV,0.5eV,0.2eV\mbox{ and } 0.1eV$ respectively.}
 \label{fig:condEfield}
\end{figure}

Now, to derive a particular type of scattering mechanism, we consider specific cases when $\tau \propto \mathcal{E}^2$, $\tau \propto \mathcal{E}$, $\tau \propto 1/\mathcal{E}$ and $\tau \sim constant$ which correspond to $\beta = +2, +1, -1, 0$ respectively. Because the electronic conductivity $\sigma$ is the only coefficient depending on frequency, we look at its various forms for the specified $\beta$ values. For $\beta = -1$; the conductivity in eq.\eqref{eq:sigma1} assumes the form
\begin{equation}
\sigma_{-1}(\Omega) = \frac{\sigma_{min}}{u_0^2}\frac{1}{1 + (\Omega\Lambda\mu)^2}.
\end{equation}
This conductivity is characterized by short-range potential that has the form contact (or delta) potential and may be due to localized impurity (defects) in the sample \cite{NMRPeres, TStauber, TAndo, BPlacias}. For $\beta = 0$; we get
\begin{equation}
 \sigma_{0}(\Omega) = \frac{\sigma_{min}}{u_0^2}\frac{\mu}{1 + (\Omega\Lambda\mu)^2},
\end{equation}
which corresponds to coherent \cite{NMRPeres} or random Dirac mass \cite{BPlacias} scattering, and describes the ballistic limit for electronic conductivity \cite{MLundstrom} in graphene. Finally, for $\beta = +1$; the electronic conductivity becomes
\begin{equation}
 \sigma_{+1}(\Omega) = \frac{\sigma_{min}}{u_0^2}\frac{\mu^2 + \dfrac{\pi^2}{3}(k_BT)^2}{1 + (\Omega\Lambda\mu)^2} \label{eq:sand1}.
\end{equation}
This is an important and dominant scattering mechanism in graphene \cite{JHChen}. It is characterized by unscreened long-range Coulomb (charged impurity) scattering \cite{SAdam}. The second term in eq.\eqref{eq:sand1} is inevitable at finite temperatures. This extra term was missing in \cite{NMRPeres}. It is the contribution due to scattering by phonons. It is clear from eq.\eqref{eq:sand1} that conductivity departs slightly from linearity behavior. Fig.\ref{fig:condEfield} depicts this situation.

The conductivity for acoustic phonon scattering is obtained at $\beta = +2$,
\begin{equation}
 \sigma_{+2}(\Omega) = \frac{\sigma_{min}}{u_0^2}\frac{\mu^3 + \pi^2(k_BT)^2\mu}{1 + (\Omega\Lambda\mu)^2}. \label{eq:sand2}
\end{equation}

\subsection*{A. Resistivity, thermal conductivity and thermopower}
In this section we turn to eqs.\eqref{eq:five} to compute other transport properties of graphene. Specifically, we will calculate the resistivity $\rho$, thermal conductivity, $\kappa$ and thermopower, $S_0$ for $\beta = +1$. We will drop the $\beta$ subscript in the following equations. By inverting eqs.\eqref{eq:five}, one can find these quantities that experimentalist usually like working with. We will write our equations similar to the format in \cite{NMRPeres}. The resistivity is,
\begin{equation}
 \rho(\Omega) = \frac{u_0^2}{\sigma_{min}}\frac{1 + (\Omega\Lambda\mu)^2}{\mu^2 + \dfrac{\pi^2}{3}(k_BT)^2}, \label{eq:rho}
\end{equation}
thermal conductivity
\begin{equation}
 \kappa(\Omega) = \frac{2}{3}\frac{\pi^2}{hT}(K_BT)^2\frac{\mu^2}{u_0^2}\, - \,\frac{8}{9}\frac{\pi^4}{hT}(K_BT)^4\frac{1 + (\Omega\Lambda\mu)^2}{1 + \dfrac{\pi^2}{3}\dfrac{(K_BT)^2}{\mu^2}}
\end{equation}
and thermoelectric power
\begin{equation}
 S_0(\Omega) = \frac{2}{3}\frac{\pi^2}{eT}\frac{(K_BT)^2}{\mu}\frac{1 + (\Omega\Lambda\mu)^2}{1 + \dfrac{\pi^2}{3}\dfrac{(K_BT)^2}{\mu^2}}.\label{eq:thermopower}
\end{equation}
The new physics emerging from these equations is the linear dependence of these quantities on $\Omega^2$. Note that electron density $n$ dependence in our equations is self manifest, since one can easily incoperate it through \cite{NMRPeres}, \cite{BPlacias} $n\propto \mu^2$ at zero temperature or 
\begin{equation}
 n =\frac{1}{\pi}\frac{1}{(\hbar v_F)^2}\left(\mu^2 + \frac{\pi^2}{3}(K_BT)^2\right)
\end{equation}
for finite temperatures.

\subsection*{IV. Magnetotransport}
The BTE for non-zero magnetic field is realized from eq.\eqref{eq:one} by making the transformation $E \to E + v \times H$ or adding the term $(v\times H)\cdot\nabla g(p)$ in the linearized BTE. This simple replacement will not yield a general solution, because of the presence of the cross product. It ensures that $v\cdot(v\times H) = 0$. To find the general solution, one usually obtains separate solutions for magnetic and electric fields and superimpose them \cite{MarkLundsrom}, \cite{JMZiman}. Here, we obtained the solution as follows; If the lattice temperature is constant of space, eq.\eqref{eq:one} becomes
\begin{equation}
 f_0(p) + \tau e'v\cdot \nabla \Phi\left(-\frac{\partial f(p)}{\partial\mathcal{E}}\right) = f(p) + \tau e(v \times  H)\cdot\nabla_pf(p). \label{eq:magneto.one}
\end{equation}
The electrochemical potential is now defined as $e\nabla \Phi = e'\nabla(\mu/e') - e'E$, with $e' = e/(1-i\Omega\Lambda\mu)$. We have assumed time-independent magnetic field in this section. The right hand side of eq.\eqref{eq:magneto.one} can be seen as an expansion of $f(p')$, where 
\begin{equation}
 p' = p+\frac{\tau ev_F}{|p|}(p \times  H) \label{eq:magneto2}
\end{equation} 
so that 
\begin{equation} 
f_0 + \tau ev\cdot \nabla \Phi\left(-\frac{\partial f}{\partial\mathcal{E}}\right) = f(p') \label{eq:magneto3}.
\end{equation}
We need to invert eq.\eqref{eq:magneto2} and put it in eq.\eqref{eq:magneto3}. To do this, we make $p$ the subject as
\begin{equation}
 p = \frac{1}{1 + \eta^2H^2}( p' - \eta p' \times  H), \label{eq:magneto4}
\end{equation} 
where $\eta = \tau e v_F/|p|$. Eq.\eqref{eq:magneto3} now becomes
\begin{equation} 
  f(p) = f_0  + \frac{\eta}{1 + \eta^2H^2} p\cdot(\nabla \Phi - \eta H\times \nabla \Phi)\left(-\frac{\partial f}{\partial\mathcal{E}}\right) \label{eq:magneto5}
\end{equation}
after dropping the prime. Notice the energy dependence of $\eta$ through $\tau$, i.e $\eta = \alpha\mathcal{E}^{\beta-1}$. Where $\alpha = \Lambda e v_F^2$ is identified as the mobility in units of $cm^2V^{-1}s^{-1}$. Now, to compute the electric current, eq.\eqref{eq:magneto5} is used in eq.\eqref{eq:three} with $\beta = +1$ to get
\begin{equation}
 J_e = \frac{\sigma(\Omega)}{1 + \alpha^2H^2}(\nabla\Phi - \alpha H\times\nabla\Phi).
\label{eq:magneto6} 
\end{equation}
The presence of magnetic field vector has created off diagonal elements in the electric current density tensor. To compute the components of the new tensor we write eq.\eqref{eq:magneto6} in an indicial notation as
\begin{equation}
 J^e_i = \frac{\sigma(\Omega)}{1 + \alpha^2H^2}(\nabla\Phi_i - \alpha \in_{ijk}H_j\nabla\Phi_k).
\label{eq:magneto7} 
\end{equation}
The longitudinal and transverse components of the magnetoconductivity tensors are 
\begin{equation}
 \sigma_{xx} = \frac{J_x}{E_x} = \frac{\sigma(\Omega)}{1 + \alpha^2H^2},
\label{eq:magnetocond}
\end{equation}
\begin{equation}
 \sigma_{xy} = \frac{J_x}{E_y} = \frac{\sigma(\Omega)\alpha H}{1 + \alpha^2H^2}.
\label{eq:magnetocond1}
\end{equation}
The rest of the components are determined through the relations $\sigma_{xx} = \sigma_{yy}$ and $\sigma_{xy} = -\sigma_{yx}$.
In terms of magnetoresistivities, eqs.\eqref{eq:magnetocond},\eqref{eq:magnetocond1} are expressed as
\begin{equation}
 \sigma_{xx} = \frac{\rho_{xx}}{\rho_{xx}^2 + \rho_{xy}^2},
\label{eq:magnetocond2}
\end{equation}
\begin{equation}
 \sigma_{xy} = \frac{\rho_{xy}}{\rho_{xx}^2 + \rho_{xy}^2}.
\label{eq:magnetocond3}
\end{equation}
with the resistivity $\rho_{xx} = 1/\sigma$ and the Hall resistivity $\rho_{xy} = \alpha H/\sigma$, giving \[\rho_{xy} = \frac{H}{ne}\] where $ne = 1/R_H$ and $R_H$ is the Hall coefficient. 

In general, $\sigma$ is complex. For this reason, we make the replacement $1/(1 + (\alpha H)^2) \to 1/(1 - i\alpha H)$ such that \[\frac{1}{1 - i\alpha H} = \frac{1 + i\alpha H}{1 + (\alpha H)^2}.\] The full electromagnetic conductivity for the $\beta=+1$ takes the form
\begin{equation}
  \sigma(\Omega) = \frac{\sigma_{min}}{u_0^2}\frac{\mu^2 + \dfrac{\pi^2}{3}(k_BT)^2}{(1-i\Omega \Lambda\mu)(1+i\alpha H)}
\end{equation}
and
\begin{equation}
  \sigma_{xy}(\Omega) = \frac{\sigma_{min}}{u_0^2}\frac{\alpha H\left(\mu^2 + \dfrac{\pi^2}{3}(k_BT)^2\right)}{(1-i\Omega \Lambda \mu)(1+i\alpha H)}
\end{equation}

\begin{figure}[h!]
 \centerline{\includegraphics[width=6.0in]{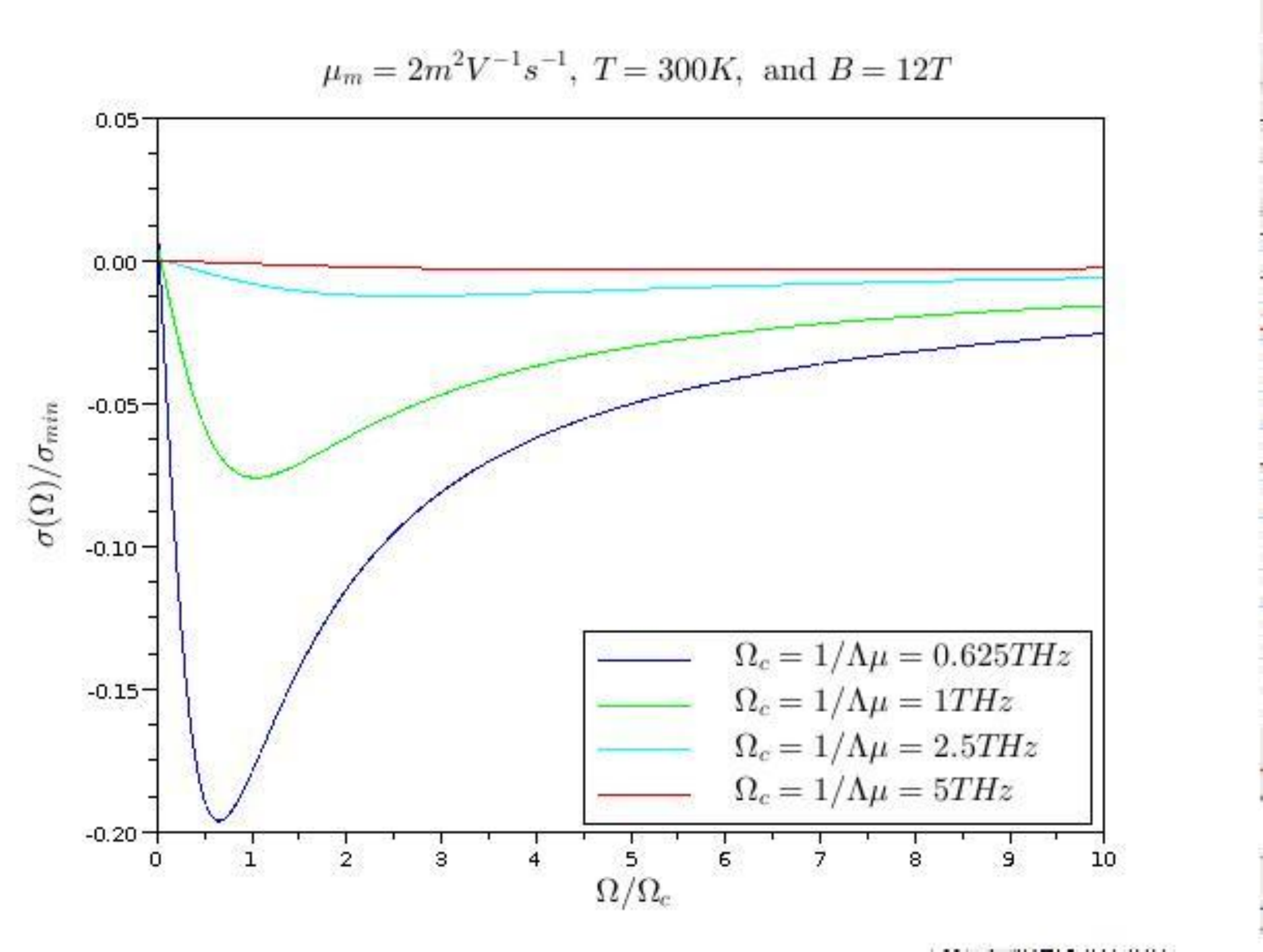}}
 \caption{Normalized conductivity is plotted against $\Omega/\Omega_c$ at fixed values of doping; $\mu = 0.8eV,0.5eV,0.2eV\mbox{ and } 0.1eV$ respectively.}
 \label{fig:condE1field}
\end{figure}

\begin{figure}[h!]
 \centerline{\includegraphics[width=6.0in]{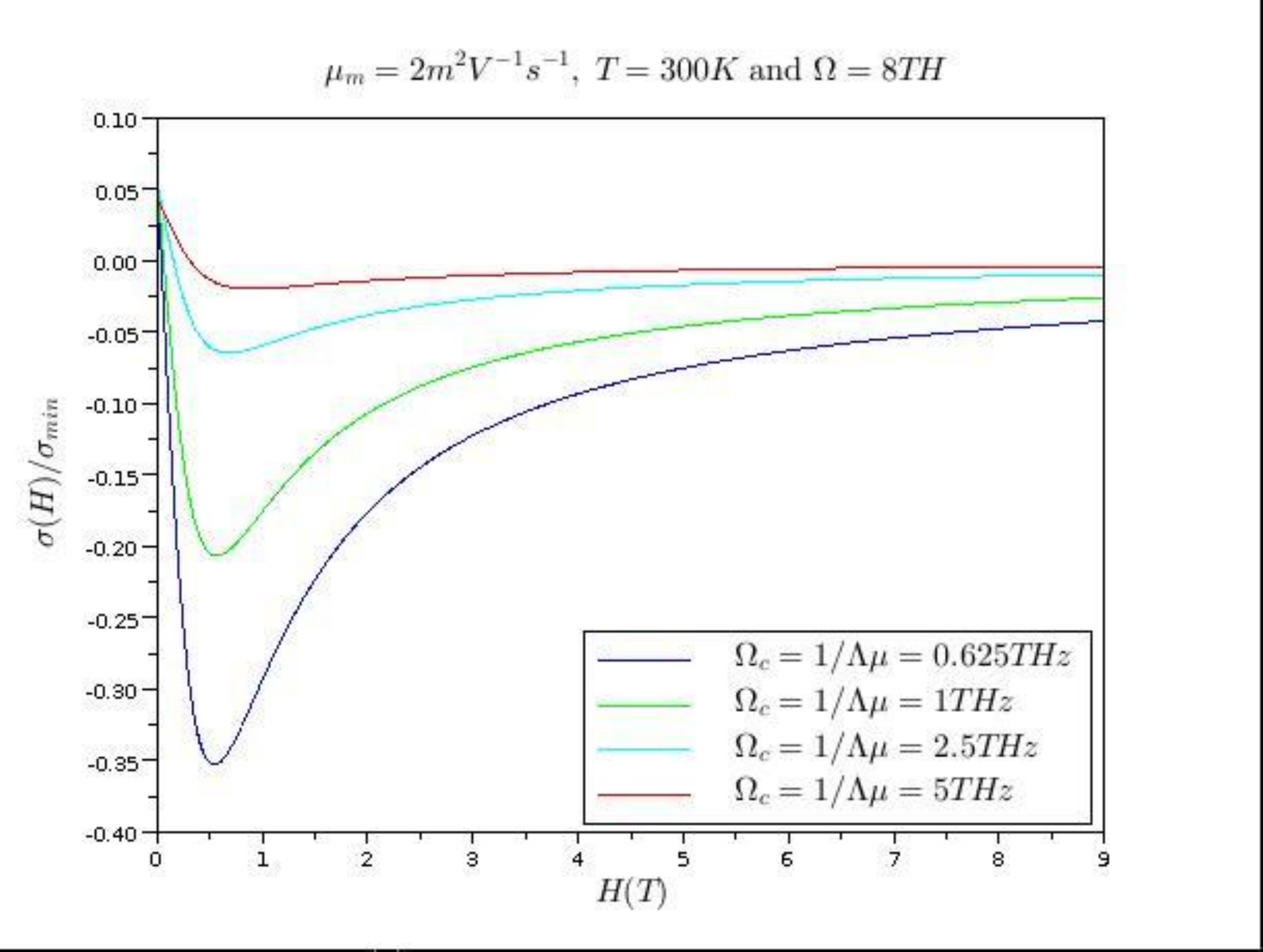}}
 \caption{Normalized conductivity is plotted against magnetic field at fixed values of doping; $\mu = 0.8eV,0.5eV,0.2eV\mbox{ and } 0.1eV$ respectively.}
 \label{fig:condBfield}
\end{figure}

We observed the effect of crossed magnetic and electric field on graphene by plotting longitudinal conductivity with electric and magnetic field in Figs.\ref{fig:condE1field},\ref{fig:condBfield}

\subsection*{V. Discussion, Conclusion}
The advantage of this approach is that, one does not need to go through rigorous process of calculating the specific scattering rate $\Gamma(\mathcal{E})$. For instance, unlike in \cite{NMRPeres} where finite temperature conductivity was found by separately calculating phonon and normal relaxation times. In this brief article, we obtained similar results without knowing a priori the exact form of the scattering potential. The challenge, however, is that certain material properties (constants), like permitivity, are not integral part of our results. Nonetheless, they can always be found by comparing with literature. But in this report, we do not care so much about numerical values of those constants, we only want to demonstrate the validity and the new physics inherent in the method. In a static electric field $\Omega = 0$, the results obtained agrees well with what is obtained in \cite{NMRPeres}, \cite{TStauber}, \cite{JNilson}, \cite{KNamura}

Using phenomenological theory based on semi-classical BTE, we reproduce transport properties of graphene without knowing the type of scattering process. We found that a characteristic exponent of $\beta = +1$ corresponds to charged impurity scattering and is a dominant mechanism in the absence of acoustic phonons. The graphs in Figs.\ref{fig:condEfield},\ref{fig:condE1field},\ref{fig:condBfield} may not be interesting. This is because of our approximations; the low energy behavior of carriers around the Dirac points. It is believed that negative differential conductivity and terahertz radiations could be observed when a full band spectrum obtained from tight-binding approximation is used \cite{SYMensah2}, \cite{SYMensah3}.

\empty


\newpage
\begin{thebibliography}{99}

\bibitem{AHCastro}
A. H. Castro Neto, \emph{et al.}, Rev. Mod. Phys. {\bf 81}, 109 (2009)

\bibitem{MIKatsenelson}
M. I Katsenelson and K. S Novoselov, ScienceDirect, Solid State Communications 143 (2007) 3 - 13

\bibitem{LCi}
L. Ci, \emph{et al.}, Nat. Mat. {\bf 9}, 430 (2010)

\bibitem{SShivaraman}
S. Shivaraman \textit{et al.}, Nano Lett. {\bf 9}, 3100 (2009)

\bibitem{FGuinea}
F. Guinea, \textit{et al.}, Phys. Rev. B {\bf 73} 245426 (2006)

\bibitem{BUchoa}
B. Uchoa, A. H. Castro Neto, \emph{Superconducting states in pure and doped graphen}, arXiv:cond-matt/0608515v3 (April 2007)

\bibitem{SAdam}
S. Adam, E. H. Hwang, V. M Galitski, and S. Dar Sarma, Proc. Nat. Aca. Sci. {\bf 104} 18392 (2007)

\bibitem{SAdam1}
S. Adam, E. H. Hwang, S. Das Sarma, physica E {\bf 40} 1022-1025 (2008)

\bibitem{AKGeim}
A. K. Geim and K. S. Novoselov , Progress Article, Nature Materials, {\bf 6} 183-191 (2007)

\bibitem{NMRPeres}
N. M. R. Peres, J. M. B. Lopes dos Santos and T. Stauber, Phys. Rev. B {\bf 76} 0743412 (2207)

\bibitem{TStauber}
T. Stauber, N. M. R. peres and F. Guinea, arXiv:0707.3004v2 (Nov 2007)

\bibitem{BPlacias}
B. Placias, \emph{Quantum Transport in Graphene}, Group de Physique Mesoscopique (MESO), (http://www.gdr-meso.phys.ens.fr) (2011)

\bibitem{JNilson}
J. Nilson \emph{et al.}, arXiv:cond-matt/0604106v2 (April 2006)

\bibitem{KNamura}
K. Namura, A. H McDonald, Phys. Rev. Lett. {\bf 96} 256602 (2006)

\bibitem{JMZiman}
J. M. Ziman, F. R. S. Melville, \emph{Principles of the theory of solids}, 2nd Edition, Cambridge University Press (1972)

\bibitem{SYMensah}
S. Y. Mensah, F. K. A. Allottey, N. G. Mensah, G. Nkrumah, J. Phys. {\bf 13} 5653 (2001)

\bibitem{MLundstrom}
M. Lundstrom, http://nanoHUB.org/topics/ElectronicsFromTheBottomUp (July, 2009)

\bibitem{MRabiu}
M. Rabiu, \emph{PhD Reserach Proposal}, University for Development Studies, Department of Applied Physics (unpublished) (2011)

\bibitem{TAndo}
T. Ando and T. Nakanishi, J. Phys. Soc. Jap.{\bf 67}, 1704, (1998)

\bibitem{JHChen}
J.H. Chen, C. Jang, S. Adam, M.S. Fuhrer, E. D. Williams, and M. Ishigami, Nature Phys., {\bf 4}, 377 (May 2008)

\bibitem{SYMensah2}
S.Y.Mensaha, A. Twuma ,N. G. Mensahb, K. A. Dompreha, S. S. Abukaria, and G. Nkrumah-Buandohc, arXiv.org/1104.1913 (2011)

\bibitem{SYMensah3}
S.Y.Mensaha, S. S. Abukaria, and N. G. Mensah, K. A. Dompreh and A. Twum, \emph{Generation of Terahertz Radiation by Wave Mixing in zigzag Carbon Nanaotubes}, arXiv.org (2011)

\bibitem{OtfriedMadelung}
O. Madelung, \textit{Introduction to Solid State Theory}, Springer-Verlag Berlin Heidelberg (1978)

\bibitem{MarkLundsrom}
Mark Lundstrom, \textit{Fundamentals of Carrier Transport}, Second Edition, (Cambridge University Press, 2000)

\bibitem{JMZiman}
J. M. Ziman, \textit{Principles of the Theory of Solids}, Second Edition, (Cambridge University Press, 1972)

\end{thebibliography}
\end{document}